\documentclass{article}
\usepackage{spconf,amsmath,graphicx,cite}
\usepackage[center]{subfigure}
\usepackage{hyperref}
\usepackage{multirow,booktabs,arydshln}
\usepackage{amssymb}

\usepackage{enumitem}
\usepackage{makecell}
\usepackage[table]{xcolor}
\definecolor{gray}{rgb}{0.85,0.85,0.85}
\usepackage{hhline}
\usepackage{tikz}
\usetikzlibrary{spy}

\usepackage{siunitx} 
\usepackage{soul}
\hypersetup{
	colorlinks   = true,    
	urlcolor     = blue,    
	linkcolor    = red,    
	citecolor    = green      
}

\newcommand*{\affmark}[1][*]{\textsuperscript{#1}}

\graphicspath{{IMAGES/}}

\title{Deep Gaussian denoiser epistemic uncertainty\\and decoupled dual-attention fusion}
\name{Xiaoqi Ma \qquad Xiaoyu Lin \qquad Majed El Helou \qquad Sabine S{\"u}sstrunk}
\address{\affmark[]School of Computer and Communication Sciences, EPFL, Lausanne, Switzerland}

\begin{document}
\maketitle
	
\begin{abstract}
Following the performance breakthrough of denoising networks, improvements have come chiefly through novel architecture designs and increased depth. While novel denoising networks were designed for real images coming from different distributions, or for specific applications, comparatively small improvement was achieved on Gaussian denoising. The denoising solutions suffer from \textit{epistemic uncertainty} that can limit further advancements. This uncertainty is traditionally mitigated through different ensemble approaches. However, such ensembles are prohibitively costly with deep networks, which are already large in size.

Our work focuses on pushing the performance limits of state-of-the-art methods on Gaussian denoising. We propose a model-agnostic approach for reducing epistemic uncertainty while using only a \textit{single pretrained network}. We achieve this by tapping into the epistemic uncertainty through augmented and frequency-manipulated images to obtain denoised images with varying error. We propose an ensemble method with two decoupled attention paths, over the pixel domain and over that of our different manipulations, to learn the final fusion. Our results significantly improve over the state-of-the-art baselines and across varying noise levels.
\end{abstract}

\noindent\let\thefootnote\relax\footnotetext{Our code, models, and supplementary material are made publicly available at: \url{https://github.com/IVRL/DEU}}

\section{Introduction} \label{sec:intro}
The importance of image denoising stems from its widespread utility in all imaging pipelines and a variety of applications. In fact, denoising can be used for regularization in general image restoration problems~\cite{cohen2020regularization}, and it is valuable when training high-level vision tasks~\cite{liu2017image}. 
Of particular interest is the fundamental problem of additive white Gaussian noise (AWGN) removal, as other noise distributions can be mapped to it with a variance stabilization transform~\cite{VST}. It has received considerable attention in the literature, where BM3D~\cite{BM3D} held for long the state-of-the-art performance among classical methods. The question of whether neural networks can compete with it~\cite{firstNNvsBM3D} was finally positively answered with the advent of deep convolutional networks for denoising~\cite{zhang2017beyond,tai2017memnet}, and the significant performance improvements that they achieved. 

\newcommand{\insetA}[1]{
\begin{tikzpicture}[baseline=0em]
    \begin{scope}[spy using outlines=
          {magnification=2, size=.8cm, connect spies}]
        \node {\includegraphics[width=0.425\linewidth]{#1}};
        \spy [red] on (0.1,1.3) in node [left] at (1.5,0.5);
    \end{scope}
\end{tikzpicture}
}

\begin{figure}[t]
	\centering
	\subfigure[Noise-free image]{
		\insetA{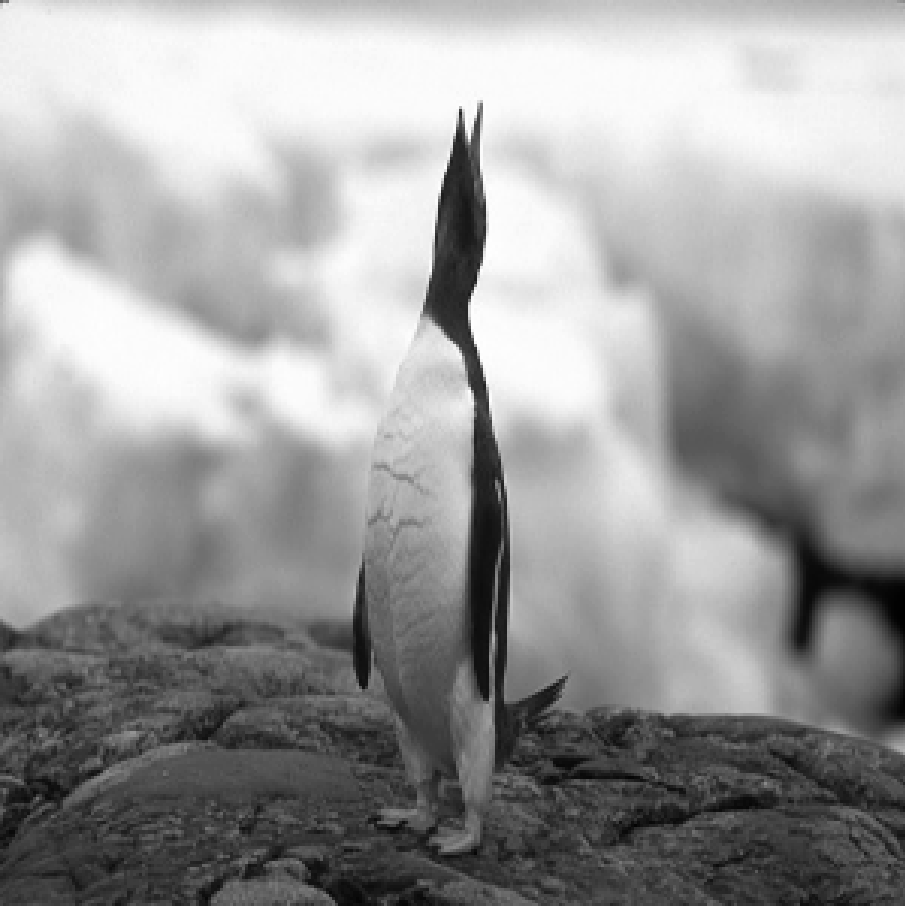}
	}
	\subfigure[Noisy image ($\sigma=50$)]{ 
		\insetA{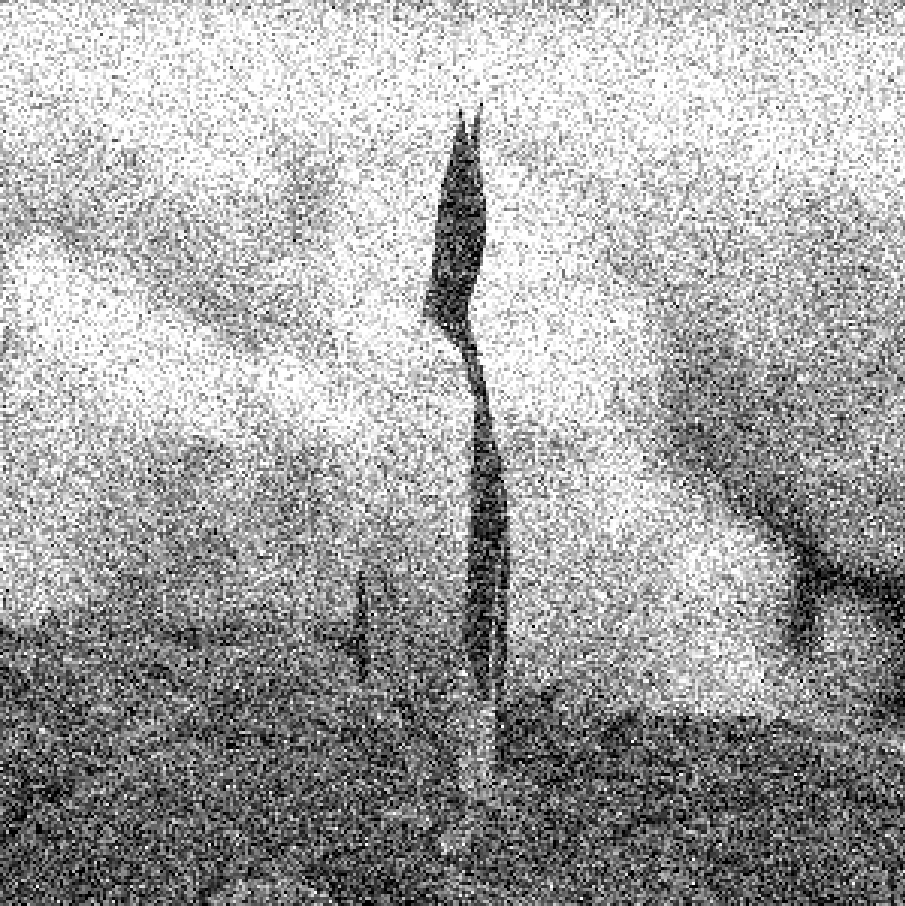}
	}\vspace{-0.25cm}\\
	\subfigure[Baseline (DnCNN)]{ 
		\insetA{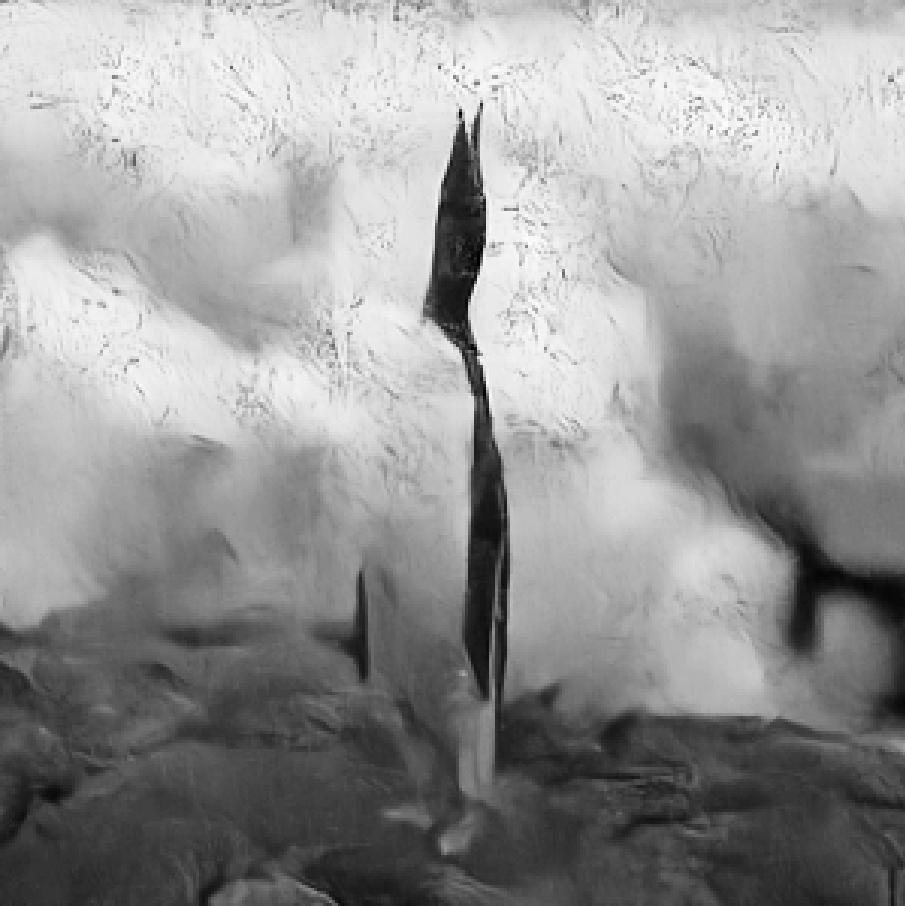}
	}
	\subfigure[With our fusion]{ 
		\insetA{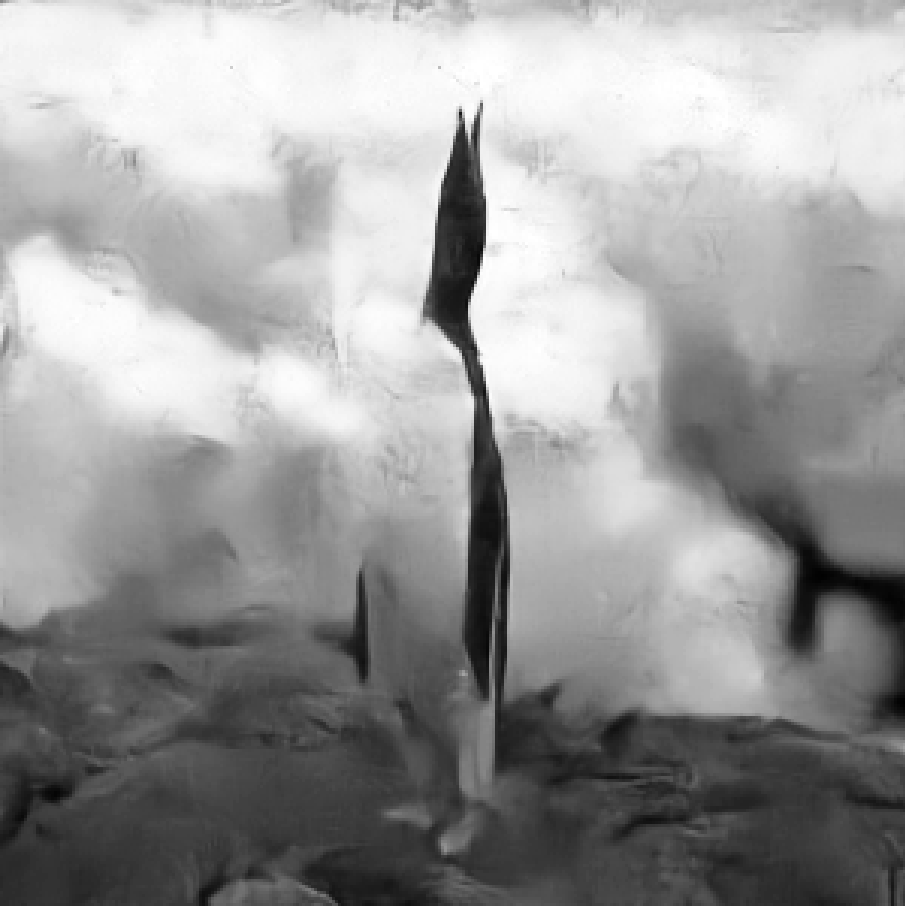}
	}
	\vspace{-0.25cm}
	\caption{Test sample denoising, showing the result of the DnCNN baseline, and our corresponding result with our attention fusion method. Best viewed on screen.}
	\label{fig:intro_images}
\end{figure}

After their initial breakthrough on AWGN removal, deep learning based denoising solutions were developed to improve their blind and universal aspects~\cite{el2020blind}, their applicability to real images~\cite{anwar2019real}, or their joint application along with demosaicking~\cite{gharbi2016deep,klatzer2016learning}, or super-resolution~\cite{zhou2020w2s}. Comparably less progress was made with respect to the performance on the AWGN removal problem, and the understanding of the performance of a given network. In~\cite{el2020blind}, the authors assess the optimality of a deep denoiser through a controlled experimental setup, and show that it does come significantly close to the statistically-optimal performance over the training range. However, the conclusions cannot be readily extended to real images because the nature of the real image prior is not analytically known~\cite{el2020blind}. A given method can hence fall short of optimality, notably due to aleatoric (or data centric) and epistemic (model centric) uncertainty. The former being mitigated in the controlled AWGN setup, we focus on epistemic uncertainty in what follows.


To address epistemic model uncertainty and improve the overall performance, ensemble methods can be leveraged.
Ensembles are made up of multiple models, with various network architectures or a fixed architecture with variable weights~\cite{collegial} obtained by retraining, or by sampling the weights from a Bayesian network~\cite{andrieu1999sequential}, or simply adding noise to the weights themselves~\cite{liu2018towards}. However, the size of the overall method grows linearly with that of the ensemble, which can be prohibitively costly with deep networks. The recent work on collegial ensembles~\cite{collegial} shows promising results that an ensemble setup can better scale compared to a wide deep network, but this setup requires a joint retraining of all the ensemble's models. In this paper, we aim to mitigate epistemic uncertainty by leveraging the power of ensembles, but using only a single pretrained network: \textit{unique architecture} and \textit{unique weights}.

To that end, we propose a self-ensembling strategy where the different branches are virtually created with a single pretrained denoiser. We empirically find that the epistemic uncertainty of a model emerges also when faced with augmented or manipulated versions of an image.
Aside from the standard spatial techniques usually used for data augmentation, we also propose frequency-domain based manipulations inspired by the training regularization masking technique recently presented in~\cite{elhelou2020stochastic}. This frequency manipulation allows us to obtain significantly less correlated denoising errors, which are crucial for any ensemble technique's performance. Attention mechanisms have shown impressive results in various applications~\cite{zhu2018end,huang2019attention,wang2020learning}, and are recently finding their way to denoising architectures~\cite{tian2020attention,li2020towards}. We make use of dual-attention paths for our ensembling method. We propose to decouple the spatial and channel attentions, leading to improved results, as we discuss in what follows.

Our results show that our method can tap into the deep denoiser epistemic uncertainty through the augmented and frequency-manipulated images, hence producing various denoised versions with variable uncertainty-based error. In other words, we virtually create an ensemble through a \textit{unique pretrained denoiser}. Furthermore, we are then able to leverage these stochastic outputs through an ensemble fusion strategy with two attention modules to significantly improve the baseline results. Our contributions can be summarized as follows. We show that the epistemic uncertainty of deep denoisers can be addressed through spatial and frequency-domain noisy input manipulations. We present a novel method to fuse the outputs that we generate, by leveraging decoupled spatial-attention and channel-attention paths. We achieve denoising improvements that are consistent across various state-of-the-art deep denoisers, and across the range of test noise levels. Our method, being denoiser agnostic, can also be applied to any novel denoising method developed in the future.

\begin{figure}
    \centering
    \includegraphics[width=\linewidth,trim={5 5 0 0},clip]{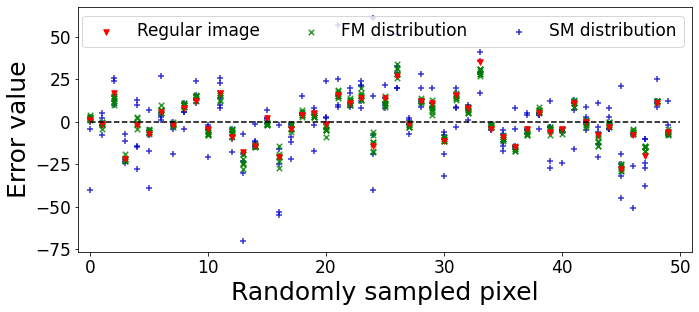}
    \caption{Denoised-image error distribution, with a pretrained DnCNN and noise level 50. Pixels are selected at random in the test set. We show the residual error in the regular denoised image, and the errors in the spatially-manipulated (SM) and frequency-manipulated (FM) images. Best viewed on screen.}
    \label{fig:error_distribution}
\end{figure}

\section{Proposed Method} \label{sec:method}
The key elements of our proposed method are the generation of a virtual ensemble using a unique pretrained network, and an ensemble fusion strategy. We discuss in this section our approach for generating the virtual ensemble, and our proposed ensemble method.

\subsection{Epistemic uncertainty through image manipulation}
We observe that, although data augmentation techniques are used during the training of deep denoising networks, the latter do not gain invariance with respect to these augmentations. For instance, the pretrained convolutional neural networks are not invariant to image mirroring, although denoising itself is agnostic to mirroring. This is one of the aspects of model uncertainty that we leverage in our method. We make use of the following seven spatial manipulations (SM): rotation of \ang{90} and vertical mirroring, vertical mirroring, rotation of \ang{270}, rotation of \ang{180} and vertical mirroring, rotation of \ang{90}, rotation of \ang{180}, and rotation of \ang{270}.

As we show in the following section, the errors across spatial manipulations remain relatively correlated. For that reason, we investigate frequency-domain image manipulations. In~\cite{elhelou2020stochastic}, the authors present a frequency-conditional learning in super-resolution networks, and by extension show that it directly relates to denoising networks as well. Based on these findings, we propose to conduct frequency manipulations (FM) by masking out different frequency bands in the noisy image. We conduct our masking essentially across the restoration target (high frequencies) but also over some of the conditional observed bands (low frequencies). The frequency-manipulated image $I_M$ is obtained as
\begin{equation}
    I_M = \mathcal{F}^{-1}( \mathcal{F}(I) \odot M),
\end{equation}
where $\mathcal{F}(\cdot)$ is a frequency transform, $I$ is the input image, $M$ is a frequency-domain mask, and $\odot$ is the element-wise product. In our work, we use the discrete cosine transform (DCT) transform type II, for its bijective relation with the Fourier transform. The mask $M$ is a binary mask delimited by quarter-annulus areas defined by two radii values. The mask is zero in the DCT domain over the quarter annulus. The radius values are computed away from the DC component of the DCT as a fraction of the maximal radius $r_{max}$. We thus make use of five frequency manipulations, corresponding to the following masks: $[0.1*r_{max},r_{max}]$, $[0.3*r_{max},r_{max}]$, $[0.5*r_{max},r_{max}]$, $[0.4*r_{max},0.5*r_{max}]$, and $[0.8*r_{max},0.9*r_{max}]$. We select the first three masks empirically to filter out what corresponds to \textit{high frequencies} relative to varying noise levels. In fact, the higher the noise level, the lower is the high-frequency restoration cutoff~\cite{elhelou2020stochastic}. The last two masks are band-stop, rather than low-pass, filters that allow the partial masking of mid-to-high frequencies. The remaining residual contributes to the variability that is beneficial for our ensemble method.

Along with the original noisy image, we thus create twelve manipulated image versions to tap into the epistemic uncertainty of a pretrained denoiser. The following section presents our ensemble method that exploits these manipulated images. We also analyse the error distribution and the correlation of error across the different proposed manipulations in Sec.~\ref{sec:analysis}.

\subsection{Decoupled dual-attention fusion}
Our ensemble method is a fusion relying on two attention mechanisms. The first attention path consists of \textit{spatial attention}, where a weight map is learned for every manipulated image. The manipulated images that are passed through the pretrained denoiser are thus element-wise multiplied with their corresponding attention maps. Different manipulations can lead to varying performance across an image. Notably, frequency manipulations can yield images with better or worse performance according to the frequency content in a given image region. Therefore, the spatial attention can learn the corresponding weight to differentiate between the varying cases. The second attention path is \textit{channel attention}. Rather than focusing on the pixel level, this attention mechanism learns to estimate the quality of the denoised images corresponding to each manipulation, as a whole. This eases the learning burden of the attention network, and provides global information on the denoising performance.

We propose to decouple the two attention paths in our fusion strategy. We note that their sequential application effectively leads to partially redundant weights. First, this redundancy reduces the overall performance of the ensemble. And second, it creates a conflict in the network's learning phase that has a negative impact on convergence. We therefore decouple our two attention paths, and merge their outputs through concatenation and a single convolutional layer. We present in our experimental results the performance of each of the attention modules separately, and show that the fusion of the two achieves the best results.

\begin{table*}[t]
	\centering
	\setulcolor{red}
	\setul{0.25ex}{0.25ex}
	\begin{tabular}{ccccccc}
		\toprule
		\makecell[c]{Backbone\\denoiser}
		& \makecell[c]{Noise\\level} 
		& \makecell[c]{Baseline\\results} 
		& \makecell[c]{Ensemble\\(SM - FM - Joint)}
		& \makecell[c]{Spatial attention\\(SM - FM - Joint)}
		& \makecell[c]{Channel attention\\(SM - FM - Joint)} 
		& \makecell[c]{Ours full\\(SM - FM - Joint)} \\ \cline{1-7}
		\noalign{\smallskip}
		
		\multirow{5}{*}{DnCNN~\cite{zhang2017beyond}}
		& 10 & \cellcolor{gray} 33.30 & 33.38 32.29 33.11 & 33.48 33.52 \ul{33.56} & 33.39 33.37 \ul{33.45} & 33.55 33.52 \textbf{33.58} \\
		& 20 & \cellcolor{gray} 29.72 & 29.78 29.25 29.67 & 29.84 \ul{29.92} 29.90 & 29.78 29.73 \ul{29.79} & 29.99 29.98 \textbf{30.03} \\
		& 30 & \cellcolor{gray} 27.68 & 27.74 27.34 27.66 & 27.83 \ul{28.02} \ul{28.02} & 27.74 27.70 \ul{27.75} & 28.12 28.10 \textbf{28.16} \\
		& 40 & \cellcolor{gray} 26.19 & 26.24 25.87 26.18 & 26.41 \ul{26.72} 26.70 & 26.24 26.25 \ul{26.29} & 26.88 26.89 \textbf{26.91} \\
		& 50 & \cellcolor{gray} 24.96 & 25.01 24.67 24.96 & 25.26 \ul{25.62} 25.55 & 25.01 25.13 \ul{25.15} & 25.96 25.97 \textbf{25.99} \\
		\noalign{\smallskip}
		
		\multirow{5}{*}{MemNet~\cite{tai2017memnet}} 
		& 10 & \cellcolor{gray} 33.40 & 33.52 32.36 33.25 & 33.52 33.55 \ul{33.60} & 33.52 33.43 \ul{33.54} & 33.64 33.52 \textbf{33.65} \\
		& 20 & \cellcolor{gray} 29.71 & 29.79 29.10 29.63 & 29.85 29.94 \ul{29.99} & 29.79 29.78 \ul{29.84} & 30.05 29.95 \textbf{30.06} \\
		& 30 & \cellcolor{gray} 27.61 & 27.68 27.10 27.55 & 27.83 28.06 \ul{28.07} & 27.68 27.77 \ul{27.81} & 28.16 28.14 \textbf{28.18} \\
		& 40 & \cellcolor{gray} 26.11 & 26.17 25.64 26.06 & 26.36 26.70 \ul{26.78} & 26.17 26.37 \ul{26.39} & 26.92 26.93 \textbf{26.94} \\
		& 50 & \cellcolor{gray} 24.94 & 24.99 24.51 24.92 & 25.22 25.62 \ul{25.80} & 24.99 \ul{25.29} 25.27 & 25.92 26.01 \textbf{26.02} \\  
		\noalign{\smallskip}
		
		\multirow{5}{*}{RIDNet~\cite{anwar2019real}}
		& 10 & \cellcolor{gray} 33.58 & 33.67 32.41 33.35 & 33.66 33.65 \ul{33.70} & \ul{33.67} 33.59 \ul{33.67} & \textbf{33.73} 33.63 \textbf{33.73} \\
		& 20 & \cellcolor{gray} 29.86 & 29.91 29.17 29.73 & 29.93 29.98 \ul{30.06} & 29.91 29.89 \ul{29.93} & 30.10 30.06 \textbf{30.11} \\
		& 30 & \cellcolor{gray} 27.71 & 27.76 27.13 27.61 & 27.87 \ul{28.11} \ul{28.11} & 27.76 27.83 \ul{27.87} & 28.22 28.19 \textbf{28.24} \\
		& 40 & \cellcolor{gray} 26.13 & 26.18 25.65 26.07 & 26.35 26.81 \ul{26.85} & 26.18 26.42 \ul{26.44} & 26.97 26.97 \textbf{27.01} \\
		& 50 & \cellcolor{gray} 24.90 & 24.95 24.50 24.88 & 25.17 \ul{25.60} 25.55 & 24.95 \ul{25.32} \ul{25.32} & 26.01 26.06 \textbf{26.08} \\
		
		\bottomrule 
	\end{tabular}
	\caption{Gaussian denoising PSNR ($dB$) results of the baseline networks, the averaging ensemble, our spatial attention module, our channel attention module, and our full dual model. We include the ablations using only our spatially-manipulated (SM) or frequency-manipulated (FM) images, rather than all (Joint). Best results in bold, best per attention mechanism are underlined.}
	\vspace{-0.15cm}

	\label{table:den_psnr_results_comparison}
\end{table*}

\begin{figure}
    \centering
    \includegraphics[width=\linewidth]{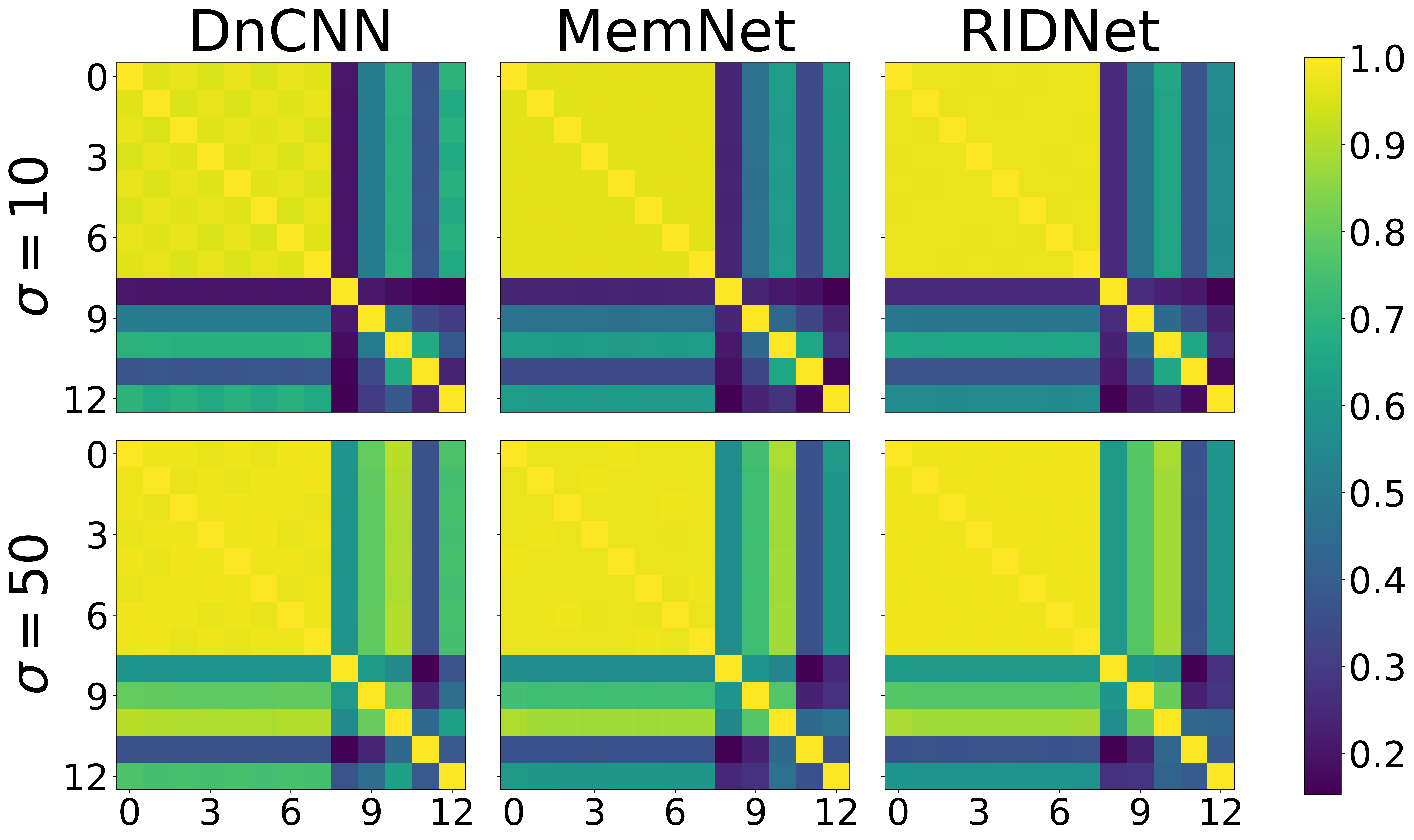}
    \caption{Pearson correlation of pixel-wise errors across the regular image and all manipulations (noise levels 10 and 50).}
    \label{fig:correlation}
\end{figure}

\newcommand{\insetB}[1]{
\begin{tikzpicture}[baseline=0em]
    \begin{scope}[spy using outlines=
          {magnification=2, size=.8cm, connect spies}]
        \node { \includegraphics[width=0.21\linewidth]{#1}};
        \spy [red] on (-0.4,0.0) in node [left] at (0.2,0.5);
    \end{scope}
\end{tikzpicture}
}

\begin{figure}{}
    \centering
    \subfigure[Noise-free\label{fig:visual_results_gt}]{
        \includegraphics[width=0.11\textwidth]{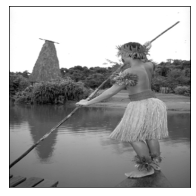}}
    \subfigure[DnCNN \label{fig:visual_results_dncnn}]{
        \includegraphics[width=0.11\textwidth]{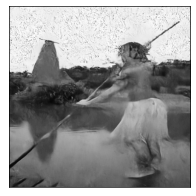}}
    \subfigure[MemNet \label{fig:visual_results_memnet}]{
        \includegraphics[width=0.11\textwidth]{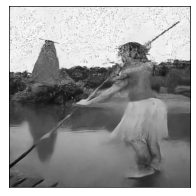}}
    \subfigure[RIDNet\label{fig:visual_results_ridnet}]{
        \includegraphics[width=0.11\textwidth]{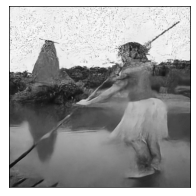}}
    \subfigure[Noisy\newline($\sigma=50$)\label{fig:visual_results_noisy}]{
        \includegraphics[width=0.11\textwidth]{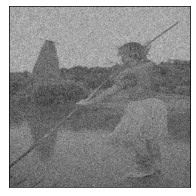}}
    \subfigure[DnCNN\newline+ Fusion \label{fig:visual_results_fusion_dncnn}]{
        \includegraphics[width=0.11\textwidth]{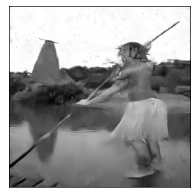}}
    \subfigure[MemNet\newline+ Fusion\label{fig:visual_results_fusion_memnet}]{
        \includegraphics[width=0.11\textwidth]{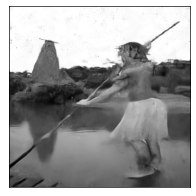}}
    \subfigure[RIDNet\newline+ Fusion\label{fig:visual_results_fusion_ridnet}]{
        \includegraphics[width=0.11\textwidth]{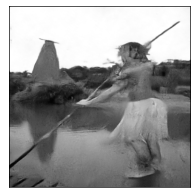}}
   
    \caption{Sample visual denoising results, with and without our proposed fusion method. Best viewed zoomed on screen.}
    \label{fig:results_images}
    \end{figure}

\section{Experimental evaluation}

\subsection{Manipulation analysis}\label{sec:analysis}
An essential component for ensembles is the variability across the underlying methods. We randomly sample from our test set 50 pixels, and analyze the error distribution in Fig.~\ref{fig:error_distribution} for different manipulations. We show the error remaining after denoising for the regular image (red triangle), the frequency-manipulated images (green crosses), and the spatially-manipulated images (blue plus symbols). We note that errors are rarely zero centered, and that the magnitude of the errors is often larger than that of the regular denoised image, hence the difficulty for the ensemble method.

We further analyze the correlation across our different manipulated and denoised images. We compute the Pearson product-moment correlation coefficients of the pixel-wise errors on the test set. These coefficients are computed pairwise, for every pair of manipulations, and the corresponding results are shown in Fig.~\ref{fig:correlation}. Index 0 corresponds to the regular denoised image, indices 1 to 7 correspond to the seven denoised images with spatial manipulations, and indices 8 to 12 to those with frequency manipulations. We first note that, although not identical, the errors across the seven spatial manipulations are significantly correlated. On the contrary, those across the five frequency manipulations are decorrelated from each other and from the spatial ones. This is a great advantage for our subsequent ensemble method, and a chief reason for proposing the frequency-masking manipulations.

\subsection{Experimental setup and results}
We conduct our experiments on the DnCNN~\cite{zhang2017beyond}, MemNet~\cite{tai2017memnet}, and RIDNet~\cite{anwar2019real} denoisers. These networks are pretrained on the BSD400 images, and are not modified or retrained in any experiment, following the experimental setup in~\cite{elhelou2020stochastic}. We train our attention-based fusion ensemble for 100 epochs on images taken from BSD500, and test the final results on the corresponding separate 100-image validation set. Our method is model agnostic and can be applied on RGB, RGBD, multi-spectral, Poisson-Gaussian, or real image denoising. Our experiments are conducted over the fundamental grayscale AWGN removal, because having multi-spectral correlated information~\cite{deblur} makes the denoising problem easier, and other noise distributions can be transformed to a normal distribution~\cite{VST}. The results are given in Table~\ref{table:den_psnr_results_comparison}. We present the results of the pretrained vanilla baseline, the straight-forward averaging (Ensemble), our method using only the spatial attention path, or only the channel attention, and our full dual-attention fusion method (Ours full). When only a single attention path is used, we fuse its different manipulated images using a softmax function for normalizing the ensemble weights. For each of the setups, we present the results when only spatial manipulations are used (SM), or only frequency manipulations (FM), or the entire set of proposed manipulations (Joint). 

The results show that, despite our manipulations that provide good error decorrelation, the averaging ensemble does not achieve any significant improvements over the baseline. In fact, the improvements with SM are slight, while with FM the results are worse than the baseline. This shows that although our frequency manipulations provide decorrelated errors, they are not zero-centered and cannot be simply averaged. We lastly note that, while the spatial or channel attention solutions can improve the final results, the best performance is consistently obtained by our decoupled dual-attention fusion. We show further visual results in Fig.~\ref{fig:results_images}, and more in our supplementary material.

\section{Conclusion} \label{sec:concl}
We present and analyze different image manipulation techniques for creating a virtual ensemble through a unique pretrained denoising network. Particularly, we obtain less correlated errors with our frequency-domain manipulations. We propose a dual attention fusion for our final ensemble, and further improve results by decoupling the attention paths. Our Gaussian denoising results consistently improve upon various denoisers, and across the test noise levels. The method we propose is denoiser agnostic and can be applied to any denoising method, and potentially other restoration tasks.

\newpage
\bibliographystyle{IEEEbib}
\bibliography{refs}

\end{document}